\documentclass[11pt]{article}

\usepackage{graphicx}
\usepackage{graphicx,psfrag}

\def \vs {\vskip 0.3cm}

\def \E {{\bf   E}}

\def \T {{\hbox{Tr\,}}}

\def \CU {{\cal U}}
\def \CN {{\cal N}}

\def \CD {{\cal D}}
\def \CI {{\cal I}}

\def \CQ {{\cal Q}}
\def \CP {{\cal P}}

\def \CD {{\cal D}}

\def\CX {{\cal X}}
\def\CY {{\cal Y}}

\def \i { {\hbox{i}}}

\def \G{{\Gamma}}

\def \a {\alpha}

\def \b {\beta}

\def \g {\gamma}

\def \l {\lambda}

\def \t {\tau}

\def \vp {\varphi}

\begin{document}

\title{On Correlation Function of  High Moments\\
of Large Wigner Random Matrices\footnote{ {\bf Acknowledgements:} The financial
support of the research grant ANR-08-BLAN-0311-11 "Grandes Matrices
Al\'eatoires" (France)  is gratefully acknowledged}\ 
}

\author{O. Khorunzhiy\\ Universit\'e de Versailles - Saint-Quentin, Versailles\\
FRANCE\\
{\it e-mail:} oleksiy.khorunzhiy@uvsq.fr}


\maketitle

\begin{abstract} 
We consider the Wigner ensemble of Hermitian $n$-dimensional
random matrices with elements $W^{(n)}_{ij}= w_{ij}/\sqrt n$ and
study the asymptotic behavior of the expression
$K_n(s',s'') = \E \left\{  \T (W^{(n)})^{2s'}\   \T (W^{(n)})^{2s''}\right\} - 
\E   \T (W^{(n)})^{2s'}\,  \E \T (W^{(n)})^{2s''}$
in the limit $n\to\infty$ such that  $s'=s'_n$ and $s''=s''_n$ are the
values of the order $n^{2/3}$.
Assuming that the  random variables  $\{w_{ij}\}$ have a symmetric
probability distribution such that its all moments $V_{2k}$ are of the sub-gaussian form,
we prove that  the limit of $K_n(s'_n,s''_n)$ exists and does not depend on the particular
values of $V_{2k}, \ k\ge 2$.

The proof is based on a combination of the arguments  by \mbox{Ya.  Sinai} and
A. Soshnikov with the  detailed study of a
moment analog of the Green's function representation of the Inverse Participation Ratio (IPR)
considered for Gaussian Unitary
Invariant Ensemble of random matrices (GUE).
\end{abstract}


\section{Introduction}

The famous semi-circle law for random matrices 
of infinitely increasing dimensions was proved by E. Wigner in the fifties by the moment method \cite{W}. 
Since that,
the moment method has been  widely used in the  spectral theory of large  random matrices. 
In particular,  it is employed in the studies of the asymptotic behavior of the matrix spectral norm 
 \cite{BY,FK,G}, where one considers the random matrix moments of the order that grows at the same time 
 as the matrix dimension
 $n$ 
 tends to infinity. In these studies, one considers the asymptotic regime when  
 the order of these moments  grows  faster 
 than $\log n$, $n\to\infty$.

 The study of the moments of the order $k \simeq n^{\alpha}$, $\alpha >0$  
can give much more detailed information on the properties of the eigenvalue distribution of random matrices
at the border of their limiting spectra.
 In a series of papers \cite{SS1,SS2,S}, Ya. Sinai and A. Soshnikov 
 have formulated 
 a deep and spacious program of the studies of the high moments of the 
 Wigner random matrices as well as the 
 correlation functions of the moments (and more generally, of their  cumulants). 
 The main aim of this program was to reach the most important  asymptotic regime when the moments of the
 order $k \simeq n^{2/3}, n\to\infty $ are investigated. A powerful approach  
 based on the original Wigner's method has been developed and the strong results of the upper bound of such moments
 as well as the universal character  of their cumulants have been stated \cite{SS1,S}.

 In papers  \cite{K3} and  \cite{KV}, we  followed this way and used the principal 
ingredients of the  method by Sinai and Soshnikov to get the rigorous proof of the upper bound
of the high moments of the Wigner  of random matrices, at the same time by relaxing the
conditions imposed in papers \cite{SS1,SS2,S} on the probability distribution of the matrix elements. 
 In the present paper we continue the work in this area 
 and  pass to the studies of the second correlation function (or the second cumulant) of these moments.

\section{Main results}

Let 
$ \{ X_{ij}, Y_{ij}, \ 1\le i\le j \}$ be  a  family of jointly independent real random variables
determined on the same probability space
and let  $\E$ denote 
the mathematical expectation with respect the  corresponding  probability
measure.
Assuming that 
$$
\E X_{ij} = \E Y_{ij} = 0, 
\eqno (2.1)
$$
and 
$$
\E X_{ij}^2 = (1+\delta_{ij})/8, \quad \E Y_{ij}^2 =
(1-\delta_{ij})/8
\eqno (2.2)
$$
for all $i$ and $j$, where $\delta_{ij}$ denotes the Kronecker $\delta$-function,
we determine complex random $n\times n$ matrices $W^{(n)}$ with the elements
$$
(W^{(n)})_{ij} = {1\over \sqrt n} 
\cases{ X_{ij} + \i Y_{ij}, 
\  \hbox{if }
1\le  i\le j\le n ,\cr
X_{ji} - \i Y_{ji},  \ \hbox{if }   1\le j<i\le n ,
\cr}
\eqno (2.3)
$$
and  say that the family $\{ W^{(n)}\}$ represents the Wigner ensemble of
Hermitian random matrices \cite{W}.
We are interested in the asymptotic behavior of  the covariance of the traces
$$
\T (W^{(n)})^{2s} = \sum_{i=1}^n (W^{2s})_{ii}
$$
in the limit when $s$ and $n$   tend to infinity. Here and below we  omit the superscripts $n$ when no confusion can arise
and say that $W= W^{(n)}$ are  the Wigner random matrices.

Our  main result is as follows.

\vskip 0.3cm   
{\bf Theorem 2.1.} {\it Let random variables $X_{ij}$ and $Y_{ij}$  have the same
symmetric probability distribution  such that, in addition to (2.2),
the moments
$
\E (X_{ij})^{2k} = V_{2k}
$
exist for all $k\ge 2$ and $V_{2k} \le (ck)^k$ for some $c>0$. Then there exists $\chi_0>0$ such that the limit }
$$
K(\chi',\chi'') = \lim_{n\to\infty} \left( \E \left\{ \T W^{2s'}\, \T W^{2s''}
\right\} - \E \,  \T W^{2s'} \  \E \,  \T W^{2s''}\right),
\eqno (2.4)
$$
{\it
where
$$
s' = s'_n = \lfloor (\chi' n^2)^{1/3}\rfloor,
\quad
s'' = s''_n = \lfloor (\chi'' n^2)^{1/3}\rfloor,
\eqno (2.5)
$$
for any given positive $\chi'$ and $\chi''$  less than $\chi_0$,
exists and does not depend
on the particular values of $V_{2k}$, $k\ge 2$. Here and below $\lfloor x\rfloor, x>0$ denotes
the maximal integer not greater than $x$.
}

\vs
Let us outline the main stages of the proof of this statement. It consists 
of two parts.  
On the first step,  the study of the covariance
$$
K_n(s',s'') =  \E \left\{ \T W^{2s'} \, \T W^{2s''} \right\} - \E  \T
W^{2s'} \, \E  \T W^{2s''}
\eqno (2.6)
$$
is reduced to the study of averages of the elements $ (W^{2m})_{ij}$ and their products 
with appropriate $m$ depending on $s'$ and $s''$.
This can be done with the help of the reduction procedure proposed in papers \cite{SS1,SS2,S}. 
In the simplest case described in these papers, this  procedure reduces the study of the expression (2.6)
to the study of the moments of the from $ \E \, \T W^{2s'+2s''-2}$. Since these moments are shown to have
an upper bound in the limit $n\to\infty, s',s'' = O(n^{2/3})$
 and the contribution of the terms that have 
the factors $V_{2k}, k\ge 2$ is shown to be vanished in this asymptotic regime, 
 a conclusion on the universality of $K_n(s',s'')$  has been stated in these works.

\vs 
However, the study of 
 $K_n(s',s'')$ (2.6)  needs not only the simple reduction procedure
considered in papers \cite{SS1,SS2,S}.
More detailed analysis  of the reduction procedures 
 shows that $K_n(s',s'')$ contains  a number of terms 
that cannot be reduced to the variables like $ \E \, \sum_{i} (W^{2s'+2s''-2})_{ii} $. These terms 
require the study of 
averages of the form 
$$
\sum_{i_1,i_2} \E \left\{ (W^{a'})_{i_1i_1} \, (W^{b'})_{i_2i_2}\right\}\  \E \left\{ (W^{a''})_{i_1i_1} \, (W^{b''})_{i_2i_2}\right\}.
\eqno (2.7)
$$
It is easy to see that (2.7) essentially differs from the moments $ \E \, \T W^{2m}$ and  combinations of them. 
From the general point of view, it is clear that the study of covariance $K_n(s',s'')$
cannot be reduced to the study of the averages $ \E \, \T W^{2m}$ only and 
represents a separate problem. 

The expressions of the form (2.7) resemble very much the well-known Green function representation 
of the Inverse Participation Ratio (IPR) that appears in the spectral theory of random matrices,
 see for example papers \cite{BK} and \cite{ESY} for the earlier and more recent references where 
 the expressions of IPR has been observed and studied, respectively. 
Therefore one could say that (2.7) represents the Moment analog of the Inverse Participation Ratio (MIPR). 

The reductions procedures leading from $K_n(s',s'')$ to $\E\, \T W^{2s'+2s''-2}$ and expression of the form (2.7)
are described in Section 3. 
The second part of the present paper (Section  5) contains the detailed analysis of the expressions of the form (2.7)
and variables related to them.
To study these variables, we propose  an  approach  based on the 
 technique developed in \cite{K2} to get the non-asymptotic estimates of   high moments 
of the random matrices of 
the 
Gaussian Unitary Invariant Ensemble (GUE) \cite{M}. The new ingredients and corresponding statements
are presented in Section 4.

For the sake of simplicity, we will say everywhere below that   $K_n(s',s'')$ (2.6)
represents the correlation function of the moments of random matrices $W^{(n)}$.
Finally, let us note that the  restrictive condition $V_{2k}\le (ck)^k$ of Theorem 2.1
is imposed  to stay close to the papers \cite{SS1} and \cite{S} and not to overload the paper. 
The upper bounds  for the high  moments of the Wigner ensemble of random matrices
are in fact proved under much more mild conditions \cite{K3}. 


\section{Correlation function and moment analog of IPR}

The study of the correlation functions of high moments of Wigner random matrices
has been started in \cite{SS1}. The reasoning of this paper
is based on the natural representation of $K_n(s',s'')$ (2.6) as a weighted sum
over the set of pairs of closed paths of $2s'$ and $2s''$ steps
$$
\CI^{(1,2)}_{2s',2s''} (n) =
\left\{ \left(I^{(1)}_{2s'}, I^{(2)}_{2s''}\right) \right\}
$$
where $I^{(1)} = (i_0^{(1)},\dots, i^{(1)}_{2s'-1}, i_0^{(1)})$,
$I^{(2)} = (i_0^{(2)},\dots, i^{(2)}_{2s''-1}, i_0^{(2)})$, and
$i^{(k)}_j \in \{1,2,\dots,n\}$ for $k=1,2$ and all $j$;
namely
$$
K_n(s',s'') = \sum_{I^{(1,2)}\in \CI^{(1,2)}_{2s',2s''} (n)}
\Pi_n( I^{(1,2)}),
\eqno (3.1)
$$
where $\Pi_n $ is a weight of the pair $I^{(1,2)}$ given by the mathematical expectation
of the product of random variables $w_{ij}/\sqrt n$ determined by $I^{(1,2)}$.

Clearly, a non-zero contribution to the right-hand side of (3.1)
is given by a subset of path pairs such that $I^{(1)}$ and $I^{(2)}$
have at least one step in common. This means that there exist two instants of time
$t'$ and $t''$ such that
$$
\left\{i^{(1)}_{t'} = i^{(2)}_{t''} \hbox{ and } i^{(1)}_{t'+1} = i^{(2)}_{t''+1}\right\}
\hbox { or }
\left\{i^{(1)}_{t'} = i^{(2)}_{t''+1} \hbox{ and } i^{(1)}_{t'+1} = i^{(2)}_{t''}\right\}
\eqno (3.2)
$$
(here and below we identify $i_{2s'}^{(1)}$ with $i_0^{(1)}$ and 
$i_{2s''}^{(2)}$ with $i_0^{(2)}$). In these two cases, we say that the step $(t',t'+1)$ of $I^{(1)}$
is passed by $I^{(2)}$ in the direct and the inverse orientations, respectively.
In what follows, we consider $t'$ and $t''$ to be the first instants of time, i.e. the minimal ones, 
with the property (3.2) to hold. The number of times that the step $(t',t'+1)$ is seen
in the path pair is referred as to the multiplicity $\mu$ of the step.

\begin{figure}
\centerline{\includegraphics[width=12cm,height=4cm]{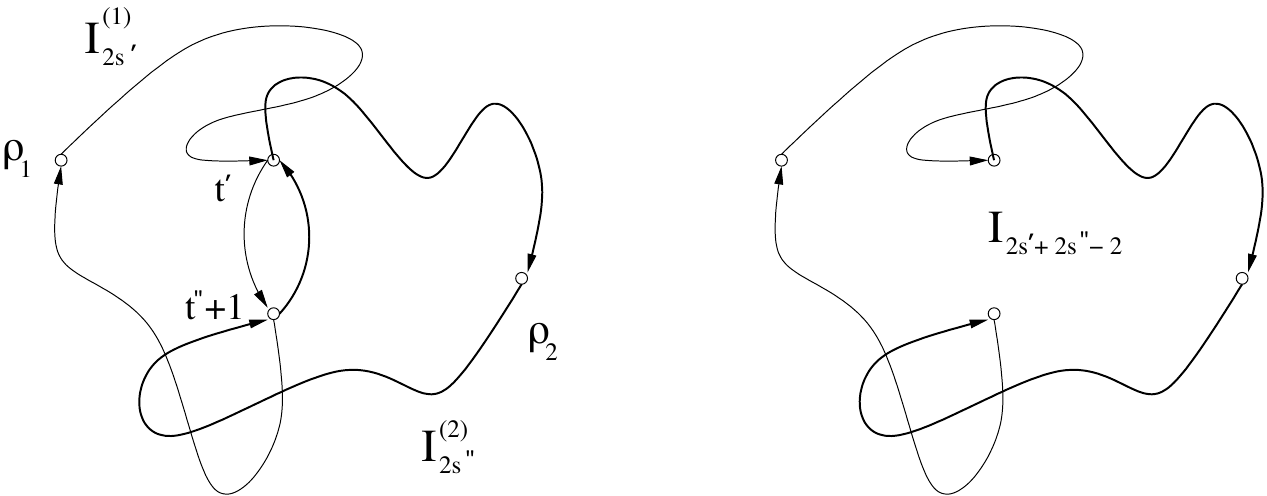}}
\caption{\footnotesize{ Simply correlated pair of paths and its two-steps reduction}}
 \end{figure}

In paper \cite{SS1}, the main attention is paid to the subset $\tilde \CI^{(1,2)}_{2s',2s''}(n)$
of path pairs such that $I^{(1)}$ and $I^{(2)}$ have at least one step
of multiplicity $\mu=2$ in common. This kind of path pairs is referred to 
as to the simply correlated pairs  \cite{SS1}.
Since $W^{(n)}$ are Hermitian, this step in common   is passed by $I^{(2)}$ in the inverse orientation
with respect to $I^{(1)}$.

By using the arguments of \cite{SS1} and \cite{S}, one can prove that
under conditions of  Theorem 2.1, the  sum
over the set of simply correlated paris
$$
\tilde \Sigma_n(s',s'') = \sum_{I^{(1,2)}\in \tilde \CI^{(1,2)}_{2s',2s''} (n)}
\Pi_n( I^{(1,2)})
\eqno (3.3)
$$
remains bounded in the limit (2.5) as $n\to\infty$ and that the contribution
of the path pairs such that there exists at least one step of multiplicity
$\mu\ge 4$ vanishes, no matter how many times this step is seen in $I^{(1)}$.

Indeed, regarding a simply correlated pair $(I^{(1)},I^{(2)})$,
one can remove the steps $(t',t'+1)$ and $(t'',t''+1)$ and consider 
the collection of remaining steps as a closed path
of $2s'+2s''-2$ steps
$$
I_{2s'+2s''-2} = (i_0^{(1)},\dots, i^{(1)}_{t'}, i^{(2)}_{t''+2},
\dots, i_0^{(2)}, i_1^{(2)},\dots, i_{t''}^{(2)}, i^{(1)}_{t'+2},\dots,
i_0^{(1)})
$$
such that
$$
\Pi (I^{(1,2)}_{2s',2s''} ) = {1\over 4n} \Pi(I_{2s'+2s''-2}).
\eqno (3.4)
$$
It is natural to say that the two-steps reduction procedure  $(I^{(1)}, I^{(2)})\to I_{2s'+2s''-2}$
is performed here.

On Figure 1 we depicted the case of simply correlated
pair and the two-step reduction procedure.
To get a non-zero weight $\Pi(I_{2s'+2s''-2})$, one has to consider
$I_{2s'+2s''-2}$ to be an even closed path \cite{SS1}.
Since
$$
\sum_{I_{2s'+2s''-2}} \Pi(I_{2s'+2s''-2}) = \E \left\{ \T (W^{(n)})^{2s'+2s''-2}\right\},
$$
one can use the result about the universal upper bound
for the averaged moments $n M^{(n)}_{2s'+2s''-2}=\E \left\{ \T (W^{(n)})^{2s'+2s''-2}\right\}$ \cite{S}
and prove the existence of their limit.  Let us stress  
that the proof of this estimate given in \cite{S} should be completed
and modified (see  \cite{K3} or \cite{KV}).
To show that $\tilde \Sigma_n(s',s'')$ remains bounded, it suffices
to estimate the number of simply correlated path pairs $I^{(1,2)}$
that can be obtained when on the base of  an even closed path $I_{2s'+2s''-2}$.
Omitting the details, one can say that this number
is proved in \cite{SS1} to be proportional to $(2s'+2s''-2)^{3/2}$ that is compensated by the factor $1/(4n)$
of (3.4)
and this
completes the argument. Let us note that the factor $(2s'+2s''-2)^{3/2}$ corresponds 
to the choice of the roots $\rho^{(1)}$ and $\rho^{(2)}$ among $2s'+2s''-2$ instants of time (see Figure 1).

\begin{figure}[htbp]
\centerline{\includegraphics[width=12cm,height=4cm]{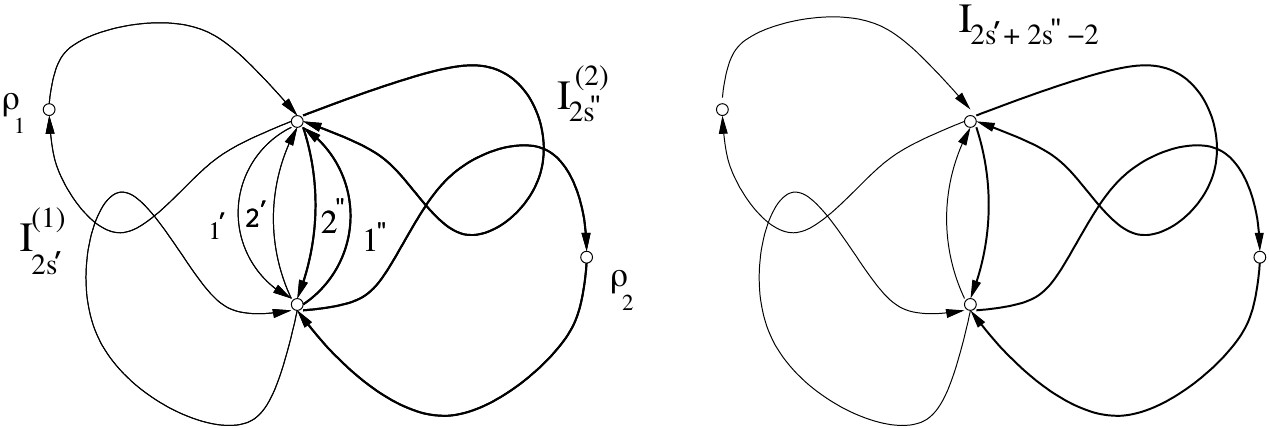}}
\caption{\footnotesize{Path pair that has  only one edge in common with $\mu=4$ and its two-steps reduction}}
 \end{figure}

As  for the path pairs that are not simply correlated ones,
it is claimed in  \cite{SS1} and subsequent papers
that the study of the corresponding sums can be also
trivially  
 reduced to the study of $M^{(n)}_{2s'+2s''-2}$.
Indeed, using the arguments described above,  it is not to hard to show
that the contribution to the right-hand side of (3.1)
that comes from the path pairs such that  $I^{(1)}$ and $I^{(2)}$ have
at least one step  of multiplicity $\mu\ge 6$ in common, vanishes in the limit (2.5) as $n\to\infty$.
The same concerns the path pairs that have one step of
multiplicity $\mu=4$ in common 
and have some other step in common passed $4$ times or more.

However, it remains one more case of non-simply correlated pairs 
that seems to be not so easy to treat.
Consider a path pair such that $I^{(1)}$ and $I^{(2)}$ have
a step of multiplicity $4$ in common and do not have any other common step.

Then the two-steps reduction procedure described above leads to  a path
$I_{2s'+2s''-2}$ that, in particular, can be either free from the steps of multiplicity
$\mu\ge 4$ or have them inside the remaining parts of $I^{(1)}$ and $I^{(2)}$
(see Figure 2).

The choice of the roots $\rho^{(1)}$ and $\rho^{(2)}$ being still 
proportional to $(2s'+2s''-2)^{3/2}$ is compensated by the factor $1/n$
and the the sum remains bounded from above. 
But it not clear how
 to prove
the fact of vanishing
contribution of the corresponding sum that we denote by 
$\hat \Sigma_n(s',s'')$
by a simple use of the two-steps reduction procedure  of \cite{SS1}.
Up to our knowledge, no rigorous study of this question has been published
or reported.
In the present paper we give a  proof
based on the method of \cite{K2}.

\vskip 0.3cm
If
the common step is passed by $I^{(1)}$ twice in opposite directions, then we get a sub-sum of 
$\hat  \Sigma_n(s',s'')$ given by expression
$$
\hat \Sigma_n^{(1)}(s',s'')=
{V_4\over n^2}\  \sum_{u,v } \sum_{i_1,i_2}
\, \E \left\{ \left[W^{2s'-2-u}_{i_1i_1}\, W^{u}_{i_2i_2} \right]^*\right\} \,
\E \left\{ \left[W^{2s''-2-v}_{i_1i_1}\, W^v_{i_2i_2}\right]^* \right\}.
\eqno (3.5)
$$
Here the symbol $[\, \cdot\,]^*$ indicates the fact that the corresponding product
does not contain the steps of multiplicity $4$ or greater.
Therefore we can bound the right-hand side of the previous inequality
by the expression
$$
R_n(s',s'') =
{V_4\over n^2} \ \sum_{u,v } \sum_{i_1,i_2}
\, \E_{\hbox {\tiny{GUE}}} \left\{ A^{2s'-u}_{i_1i_1}\, A^{u}_{i_2i_2} \right\} \,
\E_{\hbox{\tiny{GUE}}} \left\{ A^{2s''-v}_{i_1i_1}\, A^v_{i_2i_2}\right\},
\eqno(3.6)
$$
where $A=A^{(n)}$ are the random matrices of the Gaussian Unitary Invariant Ensemble
(GUE) and $\E_{\hbox {\tiny{GUE}}}$ denotes the corresponding
mathematical expectation \cite{M}. Certainly, we assume that the matrix elements
$A^{(n)}$
have the mean zero and the variance $(4n)^{-1}$ (cf. (2.2)). We have used here the fact that all terms
of the right-hand side of (3.6) are positive. 

\vskip 0.3cm
If the common step is passed by $I^{(1)}$ twice in the same direction, we get a sub-sum
$$
\hat \Sigma_n^{(2)}(s',s'')=
{V_4\over n^2}\  \sum_{u,v } \sum_{i_1,i_2}
\, \E \left\{ \left[W^{2s'-2-u}_{i_1i_2}\, W^{u}_{i_1i_2} \right]^*\right\} \,
\E \left\{ \left[W^{2s''-2-v}_{i_1i_2}\, W^v_{i_1i_2}\right]^* \right\}
$$
bounded by 
$$
S_n(s',s'') =
{V_4\over n^2} \ \sum_{u,v } \sum_{i_1,i_2}
\, \E_{\hbox {\tiny{GUE}}} \left\{ A^{2s'-u}_{i_1i_2}\, A^{u}_{i_1i_2} \right\} \,
\E_{\hbox{\tiny{GUE}}} \left\{ A^{2s''-v}_{i_1i_2}\, A^v_{i_1i_2}\right\},
\eqno(3.7)
$$
In the proof, we will see that $S_n(s',s'') = o(R_n(s',s''))$ in the limit (2.5) as $n\to\infty$. 
If the common step is passed by $I^{(1)}$ one or three times, the corresponding sub-sums
are zero. 
\vskip 0.3cm

Let us point out  that the factors and sums of  (3.6)
resemble  the expression
$$
\sum_{x=1}^n \, \E\left\{  G_{xx}(z_1)\, G_{xx}(z_2)\right\}
\eqno (3.8)
$$
where $G(z)= (A-zI)^{-1}$, $z_1 = \l + {\hbox{i}}\eta$, $z_2 = \l -\hbox{i} \eta$ with $\eta>0$,
widely used in theoretical physics in the studies of  the Inverse
Participation Ratio. This quantity  reflects the (de)localization properties
of the eigenvectors of random operators and random matrices (see paper \cite{MF} and references therein). 
Expression (3.8) and its versions naturally appear  in the studies 
of the spectral properties of random matrices 
in the asymptotic regimes when $\eta = n^{-\alpha}, \alpha >0$ \cite{BK} that is different 
from the global asymptotic regime $\eta= Const>0$.  
Recently, it has been studied
in the frameworks of the proof of the universality 
of the bulk spectral distribution of Wigner random matrices  on the microscopic scale 
when $\eta = O(1/n)$ \cite{ESY}.

It is natural to say that  (3.6) represents  a moment analog of the Green function expression (3.8). 
This becomes even clearer when one considers
the diagonal part of (3.6) (see Section 5). Therefore one can say that the main subject
of the studies of the present paper is given by the Moment representation  of the Inverse Participation Ratio
at the border of the limiting spectra. 

\section{Estimates of moments of GUE}

Let us consider the Gaussian Unitary Invariant Ensemble of random matrices
(GUE) that is given by a family of random Hermitian matrices with elements
$$
\left(A^{(n)}\right)_{xy} = {1\over \sqrt n} a_{xy}, \quad x,y =1,\dots, n
\eqno (4.1)
$$
such that the law of $A^{(n)}$ has a density
proportional to $\exp\{ -2 n\T (A^{(n)})^2\}$ \cite{M}. Then (2.1) and (2.2) hold.
For $p$ non-negative integer, we denote
$$
L^{(n)}_p = {1\over n} \sum_{x=1}^n (A^{(n)})^p_{xx} = {1\over n} \T (A^{(n)})^p,
\eqno (4.2)
$$
and consider the averaged moments
$
M_p^{(n)}= \E L^{(n)}_p,
$
where $\E $ denotes the \mbox{mathematical} expectation with respect to the probability
measure generated by $\{A^{(n)}\}$. We also denote 
$U^{(n)}_p(x) = \E\, (A^{(n)})^p_{xx}$.
Everywhere below, we will omit the superscripts $n$ when no confusion can arise.

We represent $R_n(s',s'')$ (3.6)
as a sum of four terms
$$
R_n(s',s'') = {V_4\over n^2}\  \sum_{k=1}^4 R^{(k)}_n(s',s''),
\eqno (4.3)
$$
such that
$$
R^{(1)}_n(s',s'') = \sum_{x,y=1}^n
\left( \sum_{\a_1+\b_1 = 2s'} U_{\a_1}(x)\, U_{\b_1}(y)\right)
\left(
\sum_{\a_2+\b_2 = 2s''} U_{\a_2}(x)\, U_{\b_2}(y)
\right),
$$
$$
R^{(2)}_n(s',s'') = \sum_{x,y=1}^n
\left( \sum_{\a_1+\b_1 = 2s'} U_{\a_1}(x)\, U_{\b_1}(y)\right)
\left( \sum_{\a_2+\b_2 = 2s''} \E\left\{ (A^{\a_2}_{xx})^\circ \,
(A^{\b_2}_{yy})^\circ\right\}\right),
$$
$$
R^{(3)}_n(s',s'') = \sum_{x,y=1}^n
\left( \sum_{\a_1+\b_1 = 2s'} \E\left\{ (A^{\a_1}_{xx})^\circ \,
(A^{\b_1}_{yy})^\circ\right\}\right)
\left( \sum_{\a_2+\b_2 = 2s''} U_{\a_2}(x)\, U_{\b_2}(y)\right),
$$
and
$$
R^{(4)}_n(s',s'') = \sum_{x,y=1}^n
\left( \sum_{\a_1+\b_1 = 2s'} \E\left\{ (A^{\a_1}_{xx})^\circ \,
(A^{\b_1}_{yy})^\circ\right\}\right)
\left( \sum_{\a_2+\b_2 = 2s''} \E\left\{ (A^{\a_2}_{xx})^\circ \,
(A^{\b_2}_{yy})^\circ\right\}\right),
$$
where we denoted by $X^\circ$ the centered random variable
$X^\circ = X - \E X$.
\vskip 0.5cm

In the present section we  study  the values
$U_{2s}(x)$. We do this mostly in the frameworks of the method of recurrent
non-asymptotic estimates
developed in \cite{K2} with respect to the moments $M_{2s}^{(n)}$ of GUE.
The statement we prove generalizes the results of \cite{K2} and can be formulated as follows.

\vskip 0.5cm
{\bf Theorem 4.1.} {\it Given any constant $h>1/16$, there exists
$0<\kappa<12-{3/( 4h)}$ such that the estimate
$$
\sup_{x=1,\dots,n} U_{2s}(x) \le \left( 1+ h {s(s^2-1)\over n^2}\right) m_s
\eqno (4.4)
$$
holds for all values of integer positive $s$ and $n$ satisfying condition $s^3/n^2 \le \kappa$,
where $m_s$ are the moments of the  semi-circle distribution \cite{W}
$$
m_s = {1\over 2^{2s} (s+1)}\, {{2s}\choose{s}} = {1\over 2^{2s}}
\cdot { (2s)!\over s!\, (s+1)!}, \quad s=0,1,2,\dots
\eqno (4.5)
$$
Regarding the generating function $\vp(\t) = \sum_{s=0}^\infty m_s \t^s$,
one can rewrite (4.4) in the form
$$
\sup_{x=1,\dots,n} U_{2s}(x) \le
\left[\vp(\t) +  {h \t^2\over n^2\, (1-\t)^{5/2} }\right]_s,
\eqno(4.6)
$$
where $[ \, f(\t) \,]_s= f_s$, $s=0,1,2,\dots$  denotes the coefficients of  $f(\t)=\sum_{k\ge 0} f_k \t^k$.
}

{\it Remark.}  Inequality (4.4) generalizes the non-asymptotic upper bound 
$$
M_{2s}^{(n)} \le \left( 1+ h {s(s^2-1)\over n^2}\right) m_s
$$
obtained in \cite{K2} in the same range of $s$ and $n$ as in Theorem 4.1. This bound, when considered in the limit
$s^3/n^2\to\infty$, is  asymptotically exact with the constant $h=h_0=1/16$. However, for finite values of the ratio
$s^3/n^2$, one has to increase the value of $h$ and to consider the bounded region of $\kappa$. This is because
the  exact upper bound of $M_{2s}^{(n)}$  includes the infinite series of the terms 
with the powers of $s^3/n^2$. The same reasoning explains restrictions on $\chi_0$ imposed in Theorem 2.1 (see also 
Theorem 5.1 of the next Section).

\vskip 0.5cm
The proof of Theorem 4.1 is based on the analysis of recurrent relations
for $U_{2s}(x)$ and related variables
$$
D^{(r)}_{2s}(x) =
\sum_{\a_1+\dots+\a_r=2s}
\vert \E \left\{(  A^{\a_1}  )_{xx}^\circ\,
L_{\a_2}^\circ\cdots L_{\a_r}^\circ\right\}
\vert , \quad \a_i\ge 1.
$$
In fact, we need to consider more general  than $U_{2s}(x)$  variable
$$
U_{2s}(x,y) = \E (A^{2s})_{xy} =  \sum_{t=1}^n
\, \E \left\{A_{xt}\, (A^{2s-1})_{ty}\right\}.
$$
Applying to the last mathematical expectation the integration by parts formula
(see Section 7 for the details), we get equality
$$
\sum_{t=1}^n \, \E \left\{A_{xt}\,  (A^{2s-1})_{ty}\right\} =
{1\over 4n}\  \sum_{t=1}^n \ \sum_{j=0}^{2s-2}
\E \left\{ (A^j)_{tt} \, (A^{2s-2-j})_{xy}\right\}
$$
$$
= {1\over 4 } \, \sum_{j=0}^{s-1} M_{2j}\, U_{2s-2-2j}(x,y) + \sum_{\a_1+\a_2=2s-2}
\E\left\{ (A^{\a_1})_{xy}^\circ \, L^\circ_{\a_2} \right\}.
\eqno (4.7)
$$
Introducing variable
$$
D^{(r)}_{2s}(x,y) =
\sum_{\a_1+\dots+\a_r=2s}
\vert \E \left\{(  A^{\a_1}  )_{xy}^\circ\,
L_{\a_2}^\circ\cdots L_{\a_r}^\circ\right\}
\vert ,
$$
we deduce from (4.7) the main inequality
$$
U_{2s}(x,y) \le {1\over 4} \left( M * U(x,y)\right)_{2s-2} + D^{(2)}_{2s-2}(x,y), \quad s\ge 1
\eqno (4.8)
$$
with the initial condition $ U_0(x,y) = \delta_{xy}$, $D^{(2)}_0 =0$.

Regarding the expectation
$$
\E \left\{ (A^{\a_1})_{xy} \,\left[ L_{\a_2}^\circ\cdots L^\circ_{\a_r}\right]^\circ\right\}
=\sum_{t=1}^n\E \left\{ A_{xt} \, (A^{\a_1})_{ty}\, \left[ L_{\a_2}^\circ\cdots L^\circ_{\a_r}\right]^\circ\right\}
$$
and using again the integration by parts formula, we get equality
$$
\E \left\{ (A^{\a_1})_{xy} \,\left[ L_{\a_2}^\circ\cdots L^\circ_{\a_r}\right]^\circ\right\} =
{1\over 4}\,  \sum_{j=0}^{\a_1-2}
\E\left\{ L_j (A^{\a_1-2-j})_{xy} \,
\left[ L_{\a_2}^\circ\cdots L^\circ_{\a_r}\right]^\circ\right\}
$$
$$
+
{1\over 4n^2} \
\sum_{l=2}^r \  \a_l\
\E \left\{ (A^{\a_1+\a_l-2})_{xy}\, L^\circ_{\a_2}\cdots
L^\circ_{\a_{l-1}} \, L^\circ_{\a_{l+1}}\cdots L^\circ_{\a_r}\right\}.
$$
Taking into account identity
$$
\E (XYZ^\circ) = \E X \, \E(Y^\circ Z) + \E Y \, \E(X^\circ Z)
+ \E(X^\circ  Y^\circ Z) - \E(X^\circ Y^\circ)\, \E Z
\eqno (4.9)
$$
and performing elementary transformations (see \cite{K2} for details), we get
the second main inequality
$$
D^{(r)}_{2s}(x,y) \le {1\over 4} \left( M*D^{(r)}(x,y)\right)_{2s-2}
+ {1\over 4} \left( U(x,y)*D^{(r)}\right)_{2s-2}
$$
$$
+{1\over 4} D^{(r+1)}_{2s-2}(x,y)
+{1\over 4} \left( D^{(2)}(x,y) * D^{(r-1)}\right)_{2s-2}
$$
$$
+ {r-1\over 4n^2} \left( U''(x,y) * D^{(r-2)}\right)_{2s-2}
+ {s(2s-1)(r-1)\over 4n^2}\, D^{(r-1)}_{2s-2}(x,y),
\eqno (4.10)
$$
where we denoted
$$
U''_{2s} = {(2s+2)(2s+1)\over 2}\,  U_{2s}(x,y)
$$
and $D^{(r)}_{2s} = n^{-1} \sum_{x=1}^n D^{(r)}_{2s}(x,x)$.
The initial condition for (4.10) is given by the obvious equality
$D^{(2)}_{2} (x,y) = \delta_{xy} (4n^2)^{-1}$.
Also we accept that $D^{(r)}_{2s}=0$ whenever $r>2s$ and that $D^{(1)}_{2s}=0$
and $D^{(0)}_{2s}=\delta_{s,0}$.

\vskip 0.3cm
Consider a triangular domain of integers
$$
\Delta = \{ (s,r): \ s\ge 1, \, 2\le r\le 2s\}
$$
and the numbers $\CU_s(x,y)$, $\CD^{(r)}_{2s}(x,y)$ determined in $\Delta$
by the following two systems  of recurrent relations
induced by (4.8) and 54.10), respectively,
$$
\CU_s (x,y) = {1\over 4} \, \left( \CU * \CU(x,y)\right)_{s-1} + {1\over 4}
\CD^{(2)}_{s-1}(x,y),
\eqno (4.11)
$$
and
$$
\CD^{(r)}_{s}(x,y) \le {1\over 4} \left( \CU*D^{(r)}(x,y)\right)_{s-1}
+ {1\over 4} \left( \CU(x,y)*\CD^{(r)}\right)_{s-1}
$$
$$
+{1\over 4} \CD^{(r+1)}_{s-1}(x,y)
+{1\over 4} \left( \CD^{(2)}(x,y) * \CD^{(r-1)}\right)_{s-1}
$$
$$
+ {r-1\over 4n^2} \left( \CU''(x,y) * \CD^{(r-2)}\right)_{s-1}
+ {s^2 r\over 2n^2}\, \CD^{(r-1)}_{s-1}(x,y),
\eqno (4.12)
$$
where 
$$
\CU''_s(x,y) = {(2s+2)(2s+1)\over 2} \, \CU_s(x,y), \quad \CU_s = {1\over n} \sum_{t=1}^n\,  \CU_s(t,t),
$$
$\CD^{(r)}_s =n^{-1} \sum_{t=1}^n \CD^{(r)}_s(t,t) $,
and the initial conditions are given by equalities
$$
\CU_0(x,y) = \delta_{xy}\quad \hbox{and}\quad \CD^{(2)}_1(x,y) = {\delta_{xy}\over 4n^2}.
$$
The main technical result of the present section is as follows.
\vskip 0.3cm 

{\bf Lemma 4.1.} {\it The family of numbers $\{\CU_s(x,y), \, \CD^{(r)}_s(x,y),\,(s,r)\in\Delta\}$
exists, is uniquely determined by (4.11) and (4.12)
and is such that
$$
\sup_{x,y=1,\dots, n} U_{2s}^{(n)}(x,y) \le \delta_{xy}\, \CU_s^{(n)}
\quad 
{\hbox{and}} \quad
\sup_{x,y=1,\dots,n}
D_{2s}^{(r)}(x,y)\le \CD^{(r)}_s.
\eqno (4.13)
$$
For any $h>1/12$ there exists $\kappa>0$ such that for all $1\le s\le s_0$ with $s_0^3\le \kappa n^2$
the following inequality holds
$$
\CU_s \le \left[ \vp(\t) + {h\over n^2}\, {\t^2\over (1-\t)^{5/2}}\right]_s;
\eqno (4.14)
$$
moreover, 
there exists $C$, $1/24<C< \min\{2h/3, 24\}$
such that inequalities}
$$
\CD^{(r)}_s \le 
\cases{
C (3r')!\ n^{-2r'} \left[ \t (1-\t)^{-2r'}\right]_s,  \ \hbox{if}  \ r=2r', 
\cr
C(3r'+3)! \ n^{-2r'-2} \left[ \t(1-\tau)^{-(4r'+5)/2}\right]_s,
\ \hbox{if} \     r=2r'+1
\cr}
\eqno (4.15)
$$
{\it hold for all $r$ and $s$ such that $s+2r+5\le s_0$.}

\vskip 0.3cm
The proof of Lemma 4.1 repeats almost literally  the
proof of Lemmas 2.1, 2.2 and 2.3 of \cite{K2}, so we omit the  computations
and explain only  the key points of the method. One more reason for this is that 
the main ingredients  of this method
will be used
also in the next section, where we present
the principal computations that are similar for those of the proof of Lemma 4.1.

\vskip 0.3cm 
First of all, it follows from relations (4.11) and (4.12) that $U_{2s}(x,y) = 0$
for $x\neq y$. Then we can consider the diagonal terms only
$U_{2s}(x,x)$.
Moreover, we can introduce auxiliary numbers $\bar \CU$ and $\bar\CD$
that serve as the upper bounds for $\sup_x \CU(x,x)$ and $\sup_x \CD(x,x)$
and verify relations of the form (4.11) and (4.12) and therefore satisfy (4.14)
and (4.15).

Next, assuming  $\CD^{(2)}$ to be of the order $o(1)$, it is easy to
deduce from (4.11) that the leading contribution to $\bar \CU^{(n)}_s$
is given by $[\vp(\t)]_s$. This generating function is determined as a solution
of the quadratic equation
$$
\vp(\t) = 1 + \t \vp^2(\t)/4
\eqno (4.16)
$$
and therefore is given by equality
$$
{\t \vp(\t)\over 2} = 1 - \sqrt{1-\t}.
\eqno (4.17)
$$
Assuming that $\bar \CD^{(3)}_s = o(\bar \CD^{(2)}_s)$, it is not to hard
to get from (4.12) with the help of (4.17) and the identity (see also relation (7.4) of Section 7)
$$
{(s+1)(2s+1)} m_s = \left[{1\over (1-\t)^{3/2}}\right]_s
\eqno (4.18)
$$
that
$$
\bar \CD^{(2)}_s = {C'\over n^2} {1\over (1-\tau)^2}\, (1+o(1)).
$$
Finally, regarding (4.17) and (4.18), it is natural to consider
the third derivative $\vp'''(\t)$ as a function that determines
the next-in-order corrections for $\bar \CU_s$ and therefore to
find the optimal  the form of (4.14).

Now let us return to the relation (4.11) written down for $\bar \CU_s$. 
Accepting the estimate (4.14) to hold,
we see that the first term of the right-hand side of (4.11)
can be rewritten as
\begin{eqnarray*}
{1\over 4} \left( \bar \CU* \bar \CU\right)_{s-1} &=&
\left[{\t\over 4} \left( \vp(\t) + {h\tau^2\over n^2(1-\t)^{5/2}}\right)^2\right]_s
\\
&=& \left[ \vp(\t)-1 + {h \t^2 (1-\sqrt{1-\t})\over n^2(1-\t)^{5/2}}\right]_s
+ {h^2\over 4n^2} \left[ {\t^5\over (1-\t)^5} \right]_s.
\end{eqnarray*}
The negative part of the last expression given by 
$\Phi_n(\t)=h\t^2(1-\t)^{-2} \, n^{-2}$ is exactly what we need
to compensate the contribution $C'(1-\t^2)^{-2}\, n^{-2}$ to the right-hand side of
(4.11) that comes from the estimate of $\bar \CD^{(2)}_{s-1}$. It remains to
control
the last term $h^2 [\t^5(1-\t)^{-5}]_s (4n^4)^{-1}$  to be not greater
than $[\Phi_n(\t)]_s$. This is true provided
the positive ratio $\kappa=  \sup (s^3/n^2)$ is sufficiently small.

Finally, accepting that $\bar \CD^{(5)}_s = o(\bar \CD^{(4)}_s)$, we
get from (4.12) that
$$
\bar \CD^{(4)}_s = {C''\over n^4} \left[{1\over (1-\t)^4}\right]_s (1+o(1)).
$$
This leads to the conclusion that the value
$n^{-4} [(1-\t)^{-9/2}]_s$ can be used to estimate
$\bar \CD^{(3)}_s$. The form of (4.15) for general $r$ is dictated
by (4.12) and by the detailed analysis of the expressions and constants
involved into the
computations.

\section{Correlation function terms}

In the present section we   study variables
$$
P^{(r)}_{2s}(x,y) = \sum_{\a+\b+\g_1+\dots+\g_{r-2} = 2s}
\vert \E\left\{ (A^\a)_{xx}^\circ \, (A^\b)_{yy}^\circ\,
L^\circ_{\g_1}\cdots L^\circ_{\g_{r-2}}\right\}\vert ,\quad x\neq y
$$
$$
Q^{(r)}_{2s}(x,y) = \sum_{\a+\b+\g_1+\dots+\g_{r-2} = 2s}
\vert \E\left\{ (A^\a)_{xy}^\circ \, (A^\b)_{yx}^\circ\,
L^\circ_{\g_1}\cdots L^\circ_{\g_{r-2}}\right\}\vert ,\quad x\neq y
$$
and
$$
T^{(r)}_{2s}(x) = \sum_{\a+\b+\g_1+\dots+\g_{r-2} = 2s}
\vert \E\left\{ (A^\a)_{xx}^\circ \, (A^\b)_{xx}^\circ\,
L^\circ_{\g_1}\cdots L^\circ_{\g_{r-2}}\right\}\vert ,
$$
that we refer to as to the non-crossing, crossing and diagonal terms, respectively. 

\vskip 0.3cm 

{\bf Theorem 5.1.} {\it  Under conditions of Lemma 4.1, 
there exists $\chi<\min\{ \kappa, 2^{-6}\}$
 such that the following inequalities 
}
$$
\sup_{x,y=1,\dots,n} P^{(r)}_s (x,y)\le
\cases{ 
{C\, (3r'+1)!\over n^{2r'}} \, \left[{\t\over (1-\t)^{2r'}}\right]_s,
\ \hbox{if} \   r=2r' ;\cr
{C\, (3r'+4)!\over n^{2r'+2}} \, \left[{\t\over (1-\t)^{2r'+5/2}}\right]_s,
\ \hbox{if} \   r=2r'+1,
\cr}
\eqno (5.1)
$$
\vskip 0.2cm 
$$
\sup_{x,y=1,\dots,n}  Q^{(r)}_s (x,y)\le
\cases{
 {C\, (3r')!\over n^{2r'-1}} \, \left[{\t\over (1-\t)^{2r'-3/2}}\right]_s,
\ \hbox{if}   r=2r';\cr
{C\, (3r'+3)!\over n^{2r'+1}} \, \left[{\t\over (1-\t)^{2r'+1}}\right]_s,
\ \hbox{ if} \   r=2r'+1,
\cr}
\eqno (5.2)
$$
\vskip 0.1cm 
\noindent {\it and}
$$
\sup_{x=1,\dots,n} T^{(r)}_s (x)\le
\cases{
 {C\, (3r')!\over n^{2r'-1}} \, \left[{\t\over (1-\t)^{2r'-3/2}}\right]_s,
 \ \hbox{if} \   r=2r';\cr
{C\, (3r'+3)!\over n^{2r'+1}} \, \left[{\t\over (1-\t)^{2r'+1}}\right]_s,
\ \hbox{ if} \   r=2r'+1,
\cr}
\eqno (5.3)
$$
{\it hold for all $s$ and $r$ such that $s+2r+5\le s_0$  with $s_0^3  \le \chi n^{2}$.}
\vskip 0.3cm 
We prove Theorem 5.1 by using a modification of the recurrent relations method developed in \cite{K2}.
To derive these relations, let us consider
expression
$$
H(\a,\b,\G_{r-2}) = \E\left\{ (A^\a)_{ab}^\circ \, (A^\b)^\circ_{cd}\,\, L^\circ_{\g_1}
\cdots L^\circ_{\g_{r-2}}\right\}
$$
$$
= \sum_{t=1}^n\E\left\{ A_{at}\,  (A^{\a-1})_{tb} \,
\left( (A^\b)^\circ_{cd}\, L^\circ_{\g_1}
\cdots L^\circ_{\g_{r-2}}\right)^\circ\right\}
$$
and apply to the last mathematical the integration by parts formula (7.1). We get equality
$$
H(\a,\b,\G_{r-2})
= {1\over 4 }\,  \sum_{j=0}^{\a-2}
\E\left\{  (A^{\a-2-j})_{ab} \,\,  L_j
\left( (A^\b)^\circ_{cd}\, L^\circ_{\g_1}
\cdots L^\circ_{\g_{r-2}}\right)^\circ\right\}
$$
$$
+
{1\over 4n}\, \sum_{j=0}^{\b-1}
\E\left\{ (A^j)_{ad}\,  (A^{\a+\b-2-j})_{cb} \, \, L^\circ_{\g_1}
\cdots L^\circ_{\g_{r-2}}\right\}
$$
$$
+
{1\over 4n^2}\, \sum_{l=1}^{r-2} \g_l\,
\E\left\{ (A^{\a+\g_l-2})_{ab} \, (A^\b)^0_{cd}\, \,  L_{\g_1}^\circ\cdots
L_{\g_{l-1}}^\circ \,
L_{\g_{l+1}}^\circ \cdots L_{\g_{r-2}}^\circ \right\}.
$$
Using (4.9), we obtain that $H(\a,\b,\G_{r-2})$
can be represented as a sum of ten terms,
$$
\E\left\{ (A^\a)_{ab}^\circ \, (A^\b)^\circ_{cd}\,\, L^\circ_{\g_1}
\cdots L^\circ_{\g_{r-2}}\right\}  = \sum_{k=1}^{10} J^{(k)}(\a,\b,\G_{r-2}),
\eqno (5.4)
$$
where
$$
J^{(1)}=
{1\over 4}\,  \sum_{j=0}^{\a-2} M_{j}\,  \E\left\{ (A^{\a-2-j})_{ab}^\circ \, (A^\b)^\circ_{cd}\,\,  L^\circ_{\g_1}
\cdots L^\circ_{\g_{r-2}}\right\};
$$
$$
J^{(2)} =
{1\over 4}\,  \sum_{j=0}^{\a-2} U_{\a-2-j}(a,b)\,  \E\left\{  \, (A^\b)^\circ_{cd}\,\,
L_j^\circ\, L^\circ_{\g_1}
\cdots L^\circ_{\g_{r-2}}\right\};
$$
$$
J^{(3)}= {1\over 4} \,
\sum_{j=0}^{\a-2} \E\left\{ (A^{\a-2-j})_{ab}^\circ \, (A^\b)^\circ_{cd}\,\,
L_j^\circ\, L^\circ_{\g_1}
\cdots L^\circ_{\g_{r-2}}\right\};
$$
$$
J^{(4)} = - {1\over 4} \,
\sum_{j=0}^{\a-2} \E\left\{ (A^{\a-2-j})_{ab}^\circ \, (A^\b)^\circ_{cd}\right\}
\ \E\left\{
L_j^\circ\, L^\circ_{\g_1}
\cdots L^\circ_{\g_{r-2}}\right\};
$$
$$
J^{(5)} = {1\over 4n} \, \sum_{j=0}^{\b-1}
U_{j}(a,d)\,  U_{\a+\b-2-j} (c,b) \, \E\left\{ L^\circ_{\g_1}
\cdots L^\circ_{\g_{r-2}}\right\};
$$
$$
J^{(6)} = {1\over 4n} \, \sum_{j=0}^{\b-1}
U_{\a+\b-2-j} (c,b) \, \E\left\{ (A^j)_{ad}^\circ\,\, L^\circ_{\g_1}
\cdots L^\circ_{\g_{r-2}}\right\};
$$
$$
J^{(7)} = {1\over 4n} \, \sum_{j=0}^{\b-1}
U_{j} (a,d) \, \E\left\{ (A^{\a+\b-2-j})_{cb}^\circ\,\, L^\circ_{\g_1}
\cdots L^\circ_{\g_{r-2}}\right\};
$$
$$
J^{(8)} = {1\over 4n} \, \sum_{j=0}^{\b-1}
  \E\left\{ (A^j)^\circ_{ad}\, (A^{\a+\b-2-j})_{cb}^\circ\,\, L^\circ_{\g_1}
\cdots L^\circ_{\g_{r-2}}\right\};
$$
$$
J^{(9)} = {1\over 4n^2} \, \sum_{l=1}^{r-2}
\g_l\, U_{\a+\g_l-2}(a,b) \,
\E\left\{ (A^\b)^\circ_{cd}\,\,
L_{\g_1}^\circ\cdots
L_{\g_{l-1}}^\circ \,
L_{\g_{l+1}}^\circ \cdots L_{\g_{r-2}}^\circ \right\},
$$
and
$$
J^{(10)} = {1\over 4n^2} \, \sum_{l=1}^{r-2}
\g_l\, 
\E\left\{ (A^{\a+\g_l-2})_{ab}^\circ\, (A^\b)^\circ_{cd}\,\,
L_{\g_1}^\circ\cdots
L_{\g_{l-1}}^\circ \,
L_{\g_{l+1}}^\circ \cdots L_{\g_{r-2}}^\circ \right\}.
$$

Basing on  (5.4), we derive a system of recurrent relations for
the terms $P$, $Q$ and $T$.
The structure of these relations resembles very much
the one of (4.12) and  the triangular scheme of recurrent estimates
can be applied to them. In general words, we assume the estimates
of the form (5.1), (5.2) and (5.3) to be true for all the terms of the right-hand sides
of the relations we get, and prove that the terms of the left-hand sides
also obey the corresponding estimates. More detailed discussion
of the triangular scheme of the proof of the recurrent relations
can be found in \cite{K2}.

\subsection{Relations and estimates for $P^{(r)}_{2s}(x,y)$}

Taking into account that $U_{2s}(x,y) = 0$,  $x\neq y$ and denoting
$U_{2s}(x,x) = U_{2s}(x)$, we get from (5.4)
the following system of relations for $P^{(r)}_{2s}(x,y)$,
$$
P^{(r)}_{2s}(x,y)\le {1\over 4} \left( M * P^{(r)}(x,y)\right)_{2s-2}+
{1\over 4} \left( U(x) * D^{(r)}(y)\right)_{2s-2}
$$
$$
+ {1\over 4} P^{(r+1)}_{2s-2} (x,y) +
{1\over 4} \left( P^{(2)} (x,y)* D^{(r-1)}\right)_{2s-2}
+{s\over 2n} Q^ {(r)}_{2s-2}(x,y)
$$
$$
+{r-2\over 4n^2} \left( U''(x)*D^{(r-2)}\right)_{2s-2}
+
{r-2\over 4n^2} \cdot {(2s+2)(2s+1)\over 2} \, P^{(r-1)}_{2s-2}(x,y)
\eqno (5.5)
$$
with the initial condition $P^{(2)}_2(x,y) = 0$.

\vskip 0.1cm
Let us consider the case of $P^{(2r)}_{2s}$. Regarding
the estimate of $U_{2s}(x)$ (4.6) and taking into account formula (7.5) of Section 7,
we can write that
$$
U''_{2s}(x) \le \left[ {1\over (1-\t)^{3/2}} + {18 h\over n^2}{\t^2\over
(1-\t)^{9/2}}\right]_s.
\eqno (5.6)
$$
Using these estimates and
substituting (5.1) and (5.2) to the right-hand side of (5.5), we get the
ten terms that present  as the following sum;
$$
P^{(2r)}_{2s}(x,y)\le \sum_{k=1}^{6}  \CP^{(2r;k)}_{2s},
$$
where
 \begin{eqnarray*}
 \CP^{(2r;1)}_{2s} &=& {C(3r+1)!\over 4n^{2r} }
\left[ \t^2\vp(\t)\over (1-\t)^{2r}\right]_s
\\
& = &
{1\over 2} {C(3r+1)!\over n^{2r} }\left[ \t\over (1-\t)^{2r}\right]_s
- {1\over 2} {C(3r+1)!\over n^{2r} }\left[ \t\over (1-\t)^{2r-1/2}\right]_s ;
\end{eqnarray*}

\begin{eqnarray*}
 \CP^{(2r;2)}_{2s}& =&
{C(3r)!\over 4n^{2r} }
\left[ \t^2\vp(\t)\over (1-\t)^{2r}\right]_s
\\
&=&
{1\over 2} {C(3r)!\over n^{2r} }\left[ \t\over (1-\t)^{2r}\right]_s
- {1\over 2} {C(3r)!\over n^{2r} }\left[ \t\over (1-\t)^{2r-1/2}\right]_s ;
 \end{eqnarray*}
$$
 \CP^{(2r;3)}_{2s} =
\left( {hC(3r+1)!\over 4n^{2r+2}} +{hC(3r)!\over 4n^{2r+2}} +
{C(3r+4)!\over 4n^{2r+2}}\right)
\left[\t^4\over (1-\t)^{2r+5/2}\right]_s;
$$
$$
 \CP^{(2r;4)}_{2s} =
\left( {C^2 4!(3r)!\over 4n^{2r+2}} + 
{hC(2r-2) (3r-3)!\over 4n^{2r+2}}\right)
\left[\t^3\over (1-\t)^{2r+5/2}\right]_s;
$$
$$
 \CP^{(2r;5)}_{2s} =
{s C(3r)!\over 2n^{2r}} \left[\t\over (1-\t)^{2r-3/2}\right]_s
+
{C(2r-2) (3r-3)!\over 4n^{2r}} \left[\t^4\over (1-\t)^{2r-1/2}\right]_s;
$$
and
$$
 \CP^{(2r;6)}_{2s} =
{C(2r-2)(3r+1)!\over 4n^{2r+2}}\cdot  {(2s+2)(2s+1)\over 2}
\left[\t\over (1-\t)^{2r+1/2}\right]_s.
$$
Regarding the sum of $ \CP^{(2r;1)}_{2s}$ and
$ \CP^{(2r;2)}_{2s}$, we can write that
$$
 \CP^{(2r;1)}_{2s} +  \CP^{(2r;2)}_{2s}
= {C(3r+1)!\over n^{2r}}\left[ {\t\over (1-\t)^{2r}}\right]_s
- \CX_0 - \CY_0,
\eqno (5.7)
$$
where
$$
\CX_0 = {3Cr(3r)!\over 2n^{2r} } \left[ {\t\over (1-\t)^{2r}}\right]_s
\quad {\hbox{and}} \quad 
\CY_0 =  {C(3r+2)(3r)!\over 2n^{2r} } \left[ {\t\over (1-\t)^{2r-1/2}}\right]_s.
$$
The first term of the right-hand side of (5.7) reproduces the expression
that we use as the estimate of $P^{(2r)}_{2s}$. Therefore all that we need is
to check that the sum of the remaining four terms $ \CP$ is less
than the sum $\CX_0+\CY_0$. In fact, we are going to compare
the sum of these four terms with $\CY_0$.

Using identity (7.3) (see Section 7),
we can write that for any integer  $k$
$$
\left[{\t^{1+k}\over (1-\t)^{2r+5/2}}\right]_s \le
\left[{\t\over (1-\t)^{2r+5/2}}\right]_s
$$
$$=
{ (2s+4r-3)(2s+4r-1)(2s+4r+1)\over (4r-1)(4r+1)(4r+3)}
\left[{\t\over (1-\t)^{2r-1/2}}\right]_s .
\eqno (5.8)
$$
Similar computations show that
$$
\left[{\t^4\over (1-\t)^{2r-3/2}}\right]_s \le
\left[{\t\over (1-\t)^{2r-3/2}}\right]_s
= {4r-3\over 2s+4r-5} \left[{\t\over (1-\t)^{2r-1/2}}\right]_s.
\eqno (5.9)
$$
Finally, we get equality
$$
\left[{\t\over (1-\t)^{2r+1/2}}\right]_s =
{2s+4r-3\over 4r-1} \left[{\t\over (1-\t)^{2r-1/2}}\right]_s.
\eqno (5.10)
$$
Using these three relations and taking into account that
$s+2r+1\le s_0 = \chi n^{2/3}$, we can write that
$$
{1\over \CY_0} \sum_{k=3}^{6}  \CP^{(2r;k)}_{2s} \le
{16h\chi\over (4r-1)^3} + {4\chi} {(3r+1)(3r+3)(3r+4)\over (4r-1)(4r+1)(4r+3)} +
{192C\chi\over(3r+2)(4r-1)(4r+3)}
$$
$$
+ {(4r-3)s\over (3r+2)(2s+4r-5)}
+ {r-1\over (3r-2)(3r-1)3r(3r+2)} + {18 h\chi\over (4r-1)(4r+1)(4r+3)}. 
$$
Taking into account that $r\ge 1$, we see that the right-hand side of
the last inequality is strictly less than $5/6$ when $h$ and $\chi$
verify  conditions of Theorem 5.1. The estimate (5.1) is proved for
$P^{(2r)}_{2s}$.

\vskip 0.3cm

Let us consider relation (5.5) for $P^{(2r+1)}_{2s}$. Taking into account
estimates  (4.6) and (5.6) and using expression (4.17),
we get an inequality for $ P^{(2r+1)}_{2s}$ that counts ten terms that we present
as the following sum
$$
P^{(2r+1)}_{2s}(x,y) \le \sum_{k=1}^{6}
 \CP^{(2r+1;k)}_{2s},
\eqno (5.10)
$$
where
 \begin{eqnarray*}
 \CP^{(2r+1;1)}_{2s} 
&=& {C(3r+4)!\over 4n^{2r+2}}
\left[{\t^2\vp(\t)\over (1-\t)^{2r+5/2}}\right]_s
\\
&=&{C(3r+4)!\over 2n^{2r+2}}
\left[{\t\over (1-\t)^{2r+5/2}}\right]_s - {C(3r+4)!\over 2n^{2r+2}}
\left[{\t\over (1-\t)^{2r+2}}\right]_s;
 \end{eqnarray*}
 \begin{eqnarray*}
 \CP^{(2r+1;2)}_{2s} &=&
 {C(3r+3)!\over 4n^{2r+2}}
\left[{\t^2\vp(\t)\over (1-\t)^{2r+5/2}}\right]_s
\\
&=&{C(3r+3)!\over 2n^{2r+2}}
\left[{\t\over (1-\t)^{2r+5/2}}\right]_s - {C(3r+3)!\over 2n^{2r+2}}
\left[{\t\over (1-\t)^{2r+2}}\right]_s;
\end{eqnarray*}
 $$
 \CP^{(2r+1;3)}_{2s} = \left( 
{9Ch(3r+4)!\over 2n^{2r+4}} + {9Ch(3r+3)!\over 2n^{2r+4}} +
 {9hC(2r-1) (3r)!\over 2n^{2r+4}}\right)
\left[{\t^4\over (1-\t)^{2r+5}}\right]_s;
$$
$$
 \CP^{(2r+1;4)}_{2s} =\left(  {C(3r+4)!\over 4n^{2r+2}}
+{C^24!(3r)!\over 4n^{2r+2}}\right)
\left[{\t\over (1-\t)^{2r+2}}\right]_s;
$$
$$
 \CP^{(2r+1;5)}_{2s} = {sC(3r+3)!\over 2n^{2r+2}}
\left[{\t\over (1-\t)^{2r+1}}\right]_s + {C(2r-1)(3r)!\over 4n^{2r+2}}
\left[{\t^2\over (1-\t)^{2r+2}}\right]_s;
$$
and
$$
 \CP^{(2r+1;6)}_{2s} = {C(2r-1) (3r+1)!\over 2n^{2r+4}}
\cdot {(2s+2)(2s+1)\over 2}
\left[{\t\over (1-\t)^{2r}}\right]_s.
$$

Regarding the sum of two first terms, we can write that
$$
 \CP^{(2r+1;1)}_{2s} +  \CP^{(2r+1;2)}_{2s} =
{C(3r+4)!\over 2n^{2r+2}}
\left[{\t\over (1-\t)^{2r+5/2}}\right]_s -\CX_1 - \CY_1,
$$
where
$$\CX_1 = {C(3r+5)\, (3r+3)!\over 2n^{2r+2}}
\left[{\t\over (1-\t)^{2r+2}}\right]_s
$$
and
$$
\CY_1 = {C(3r+3)\, (3r+3)!\over 2n^{2r+2}}
\left[{\t\over (1-\t)^{2r+5/2}}\right]_s.
$$
Using identity  (7.3) of Section 7, it is not hard to show that for any integer
$s\ge 1$
$$
\left[{\t\over (1-\t)^{2r+5/2}}\right]_s\ge \left[{\t\over (1-\t)^{2r+2}}\right]_s.
$$
Then we can write that
$$
\CX_1 + \CY_1 \le {C(3r+3)\, (3r+3)!\over n^{2r+2}}
\left[{\t\over (1-\t)^{2r+2}}\right]_s = \tilde \CX_1.
$$
Now it remains to show that the sum
$\sum_{k=3}^{10}  \CP^{(2r+1;k)}_{2s}$ does not exceed
$\tilde \CX_1$.
To do this, we will use (7.6) and its consequences
given by the following three relations;
$$
\left[{\t^4}\over (1-\t)^{2r+5}\right]_s \le {s^3\over (2r+2)(2r+3)(2r+4)}
\left[{\t}\over (1-\t)^{2r+2}\right]_s,
\eqno (5.11)
$$
$$
\left[{\t}\over (1-\t)^{2r+1}\right]_s =
{2r+1\over s+2r}\left[{\t}\over (1-\t)^{2r+2}\right]_s,
\eqno (5.12)
$$
and
$$
\left[{\t}\over (1-\t)^{2r}\right]_s = {2r(2r+1)\over (s+2r-1)(s+2r)}
\left[{\t}\over (1-\t)^{2r+2}\right]_s.
\eqno (5.13)
$$

Using these three relations, we can write that
$$
{1\over \tilde \CX_1} \sum_{k=3}^{6}  \CP^{(2r+1;k)}_{2s} \le
{9h\chi (3r+5)\over 4(3r+3)(2r+2)(2r+3)(2r+4)}
$$
$$
+ {3r+4\over 4(3r+5)} + {3C\over (3r+1)(3r+2)(3r+3)^2}
+{2r+1\over 2(3r+5)} + {r\over (3r+1)(3r+2)(3r+3)^2}
$$
$$
+ {9h\chi (2r-1)\over 2(3r+1)(3r+2)(3r+3)^2} + {r(2r-1)(2r+1)\over (3r+2)(3r+3)^2}.
$$
It is clear that under conditions of Theorem 5.1 the right-hand side
of the last inequality
is strictly less than $1$ for any $r\ge 1$. The estimate (5.1)
is proved for the variable $P_{2s}^{(2r+1)}(x,y)$.

\subsection{Relations and estimates for $Q^{(r)}_{2s}(x,y)$}

Regarding (5.4) with $a=d=x$ and $b=c=y$ and taking into account equality
$U_{2s}(x,y)=0$ for $x\neq y$, we obtain the following recurrence;
$$
Q^{(r)}_{2s}(x,y)\le {1\over 4} \left( M*Q^{(r)}(x,y)\right)_{2s-2} +
{1\over 4} Q^{(r+1)}_{2s-2}(x,y)
$$
$$
+{1\over 4} \left(Q^{(2)}_{2s-2}(x,y)*D^{(r-1)}\right)_{2s-2}
+{1\over 4n} \left(U'(x)*U(y)*D^{(r-2)}\right)_{2s-2}
$$
$$
+{1\over 4n} \left(U'(y)*D^{(r-1)}(x)\right)_{2s-2}
+ {s\over 2n} \left(U(x)*D^{(r-1)}(y)\right)_{2s-2}
$$
$$
+ {s\over 2n} P^{(r)}_{2s-2}(x,y)
+{(r-2)\over 4n^2} \cdot {(2s+2)(2s+1)\over 2} \, Q^{(r-1)}_{2s-2}(x,y)
\eqno (5.14)
$$
with the initial condition $Q^{(2)}_2(x,y) = 1/(4n)$,
where we denoted
$$
U'_{2s}(x) = (2s+2) U_{2s}.
$$

Let us consider the case of $Q^{(2r)}_{2s}$. Regarding (4.6) and using formulas (7.4),
we obtain that
$$
U'_{2s}(x)\le 2 \left[ {1\over \sqrt{1-\t}} +
{h\t^2\over n^2\, (1-\t)^{7/5}}\right]_{s}.
\eqno (5.15)
$$
Using this estimate and substituting (5.1) and (5.2) to the right-hand side of
(5.14), we get twelve terms that we regroup in the sum of six terms
as follows;
$$
Q^{(2r)}_{2s}(x,y) \le \sum_{k=1}^{6} \CQ^{(2r;k)}_{2s},
\eqno (5.16)
$$
where
\begin{eqnarray*}
 \CQ^{(2r;1)}_{2s} &=& {C(3r)!\over 4n^{2r-1}}
\left[{\t^2\vp(\t)\over (1-\t)^{2r-3/2}}\right]_s
\\
&=& {C(3r)!\over 2n^{2r-1}}
\left[{\t\over (1-\t)^{2r-3/2}}\right]_s - {C(3r)!\over 2n^{2r-1}}
\left[{\t\over (1-\t)^{2r-2}}\right]_s;
\end{eqnarray*}
 $$
\CQ^{(2r;2)}_{2s} = {C\over 4n^{2r+1}}
\left(h(3r)!+ (3r+3)!+ C 3!(3r)!+ 2h(3r-3)!\right)
\left[{\t^2\over (1-\t)^{2r+1}}\right]_s;
$$
$$
\CQ^{(2r;3)}_{2s}  = {C(3r)!\over 2n^{2r+1}} \left(
s \left[ {\t^2\over (1-\t)^{2r}}\right]_s
+
(r-1)(s+1)(2s-1) \left[ {\t\over (1-\t)^{2r-1}}\right]_s\right);
$$
$$
\CQ^{(2r;4)}_{2s}  = {Ch\over 2n^{2r+3}}
\left( h (3r-3)!\left[ {\t^6\over (1-\t)^{2r+4}}\right]_s
+ s(3r)! \left[ {\t^4\over (1-\t)^{2r+3}}\right]_s\right);
$$
$$
\CQ^{(2r;5)}_{2s} = {C(3r-3)!\over  2n^{2r-1} }
\left[ {\t^4\over (1-\t)^{2r-3/2}}\right]_s,
$$
and
$$
\CQ^{(2r;6)}_{2s} = {C\over  2n^{2r+1} }
\left(
s(3r)!\left[ {\t^2\over (1-\t)^{2r+1/2}}\right]_s
+ h (3r-3)!\left[ {\t^4\over (1-\t)^{2r+3/2}}\right]_s\right).
$$

Regarding $\CQ^{(2r;1)}_{2s}$, we can write that
$$
\CQ^{(2r;1)}_{2s} =
{C(3r)!\over n^{2r-1}}
\left[{\t\over (1-\t)^{2r-3/2}}\right]_s
- \CX_2 - \CY_2,
\eqno (5.17)
$$
where
$$
\CX_2 =  {C(3r)!\over 2n^{2r-1}}
\left[{\t\over (1-\t)^{2r-2}}\right]_s,
\quad
\CY_2 =  {C(3r)!\over 2n^{2r-1}}
\left[{\t\over (1-\t)^{2r-3/2}}\right]_s.
$$
We are going to show that the sum
$
\sum_{k=2,3,4} \CQ^{(2r;k)}_{2s} $ is strictly less than
$\CX_2$ and that the sum
$ \CQ^{(2r;5)}_{2s}+ \CQ^{(2r;6)}_{2s}$ is strictly less than
$\CY_2$.
To do this, we will use the following consequences of
formulas (7.3) and (7.6):
$$
\left[ {\t\over (1-\t)^{2r-2 +q}}\right]_s
= {(s+2r-3)(s+2r-2)\cdots (s+2r-4+q)\over
(2r-1) 2r \cdots (2r-2+q)}\,
\left[ {\t\over (1-\t)^{2r-2}}\right]_s,
$$
for $ q=1,2,3,5,6$
and
$$
\left[ {\t\over (1-\t)^{2r+1/2 }}\right]_s
= {(2s+4r-5)(2s+4r-3)\over (4r-3)(4r-1)}\left[ {\t\over (1-\t)^{2r-3/2}}\right]_s
\eqno (5.18)
$$
and
$$
\left[ {\t\over (1-\t)^{2r+3/2 }}\right]_s
= {(2s+4r-5)(2s+4r-3)(2s+4r-1)\over (4r-3)(4r-1)(4r+1)}
\left[ {\t\over (1-\t)^{2r-3/2}}\right]_s.
\eqno (5.19)
$$
Using (5.17); we conclude that
$$
{1\over \CX_2} \sum_{k=2,3,4} \CQ^{(2r;k)}_{2s} \le
\chi {(3r+1)(3r+2)(3r+3) + h +3C \over 2r (2r-1) (2r+1)}
$$
$$
+ {\chi\over 2r(2r-1)} + {\chi}
+
{h\chi^2 \over 7!} + {h^2\chi^2\over 8!},
\eqno (5.20)
$$
where the right-hand side is strictly less than $1$
for any $s,r\ge 1$ and $s_0^3/n^2\le \chi$.
Also we have inequality
$$
{1\over \CY_2}(\CQ^{(2r;5)}_{2s}+ \CQ^{(2r;6)}_{2s})\le
{1\over 3r(3r-1)(3r-2)}
+ {\chi\over (4r-1)(4r-3)}
$$
$$
+ {h\chi\over (4r-1)(4r+1)(3r)(3r-1)(3r-2)},
\eqno (5.21)
$$
where the right-hand side is strictly less than $1$ under conditions of Theorem 5.1.

\vskip 0.3cm
Let consider $Q^{(2r+1)}_{2s}(x,y)$. We have from (5.14) that
$$
Q^{(2r+1)}_{2s}(x,y) \le \sum_{k=1}^{6} \CQ^{(2r+1;k)}_{2s},
\eqno (5.22)
$$
where
\begin{eqnarray*}
 \CQ^{(2r+1;1)}_{2s} &=&
{C(3r+3)!\over 4n^{2r+1}}
\left[{\t^2\vp(\t) \over (1-\t)^{2r+1}}\right]_s
\\
&=&
{C(3r+3)!\over 2n^{2r+1}}
\left[{\t \over (1-\t)^{2r+1}}\right]_s
-
{C(3r+3)!\over 2n^{2r+1}}
\left[{\t \over (1-\t)^{2r+1/2}}\right]_s;
\end{eqnarray*}
 $$
\CQ^{(2r+1;2)}_{2s} = {C\over 4n^{2r+1}} ((3r+3)! + 6C(3r)!/n)
\left[{\t \over (1-\t)^{2r+1/2}}\right]_s
$$
$$
+
C(3r)!{(2r-1)(s+1)(2s+1)\over 4n^{2r+1} }
\left[{\t \over (1-\t)^{2r-3/2}}\right]_s;
$$
$$
\CQ^{(2r+1;3)}_{2s} = {sC\over 2n^{2r+3}}
(h(3r)! + (3r+3)!)\, \left[{\t \over (1-\t)^{2r+5/2}}\right]_s;
$$
$$
\CQ^{(2r+1;4)}_{2s} = {hC\over 2n^{2r+3}} (C(3r+3)! + (3r)!)
\left[{\t \over (1-\t)^{2r+7/2}}\right]_s
+
{h^2C(3r)!\over 2n^{2r+4}} \left[{\t \over (1-\t)^{2r+13/2}}\right]_s
$$
and
$$
\CQ^{(2r+1;5)}_{2s} = {C(3r)!\over 2n^{2r+1}}
\left( s \left[{\t \over (1-\t)^{2r}}\right]_s
+2 \left[{\t \over (1-\t)^{2r+1}}\right]_s
+ {h\over n^2}\left[{\t \over (1-\t)^{2r+4}}\right]_s \right).
$$

Regarding $ \CQ^{(2r+1;1)}_{2s} $, we can write that
$$
\CQ^{(2r+1;1)}_{2s} =
{C(3r+3)!\over n^{2r+1}}
\left[{\t \over (1-\t)^{2r+1}}\right]_s
- \CX_3 - \CY_3,
\eqno (5.23)
$$
where
$$
\CX_3 = {C(3r+3)!\over 2n^{2r+1}}
\left[{\t \over (1-\t)^{2r+1}}\right]_s,\quad
\CY_3 = {C(3r+3)!\over 2n^{2r+1}}
\left[{\t \over (1-\t)^{2r+1/2}}\right]_s;
$$
Now it is sufficient to show that
$\sum_{k=2,3,4}  \CQ^{(2r+1;k)}_{2s}$ is strictly less than $\CY_3$
and that
$ \CQ^{(2r+1;5)}_{2s}$ is strictly less than $\CX_3$.

\vskip 0.5cm
It follows from (7.6) that 
$$
\left[ {\t\over (1-\t)^{2r-3/2} } \right]_s = {( 4r-3)(4r-1)\over (2s+4r-3)(2s+4r-1)} 
\left[ {\t\over (1-\t)^{2r+1/2} } \right]_s
$$
and that 
$$
\left[ {\t\over (1-\t)^{2r+1/2+q} } \right]_s\le { (2s_0)^q\over (4r)^q} \left[ {\t\over (1-\t)^{2r+1/2} } \right]_s.
$$
Then we easily get inequality
$$
{1\over \CY_3} \sum_{k=2,3,4} \CQ^{(2r+1;k)}_{2s} \le {1\over 2} + {8\over 27} + {h\chi (1+C)\over 16}  + 
{h^2\chi^3\over 4^9}.
\eqno (5.24)
$$
Clearly, the right-hand side of it is less than 1 under conditions of Theorem 5.1.

\vskip 0.5cm

Using (7.6), we can write that
$$
{1\over \CX_3} \CQ^{(2r+1;5)}_{2s} \le {2rs\over (s+2r) (3r)^3} + {2\over (3r)^3 } + {h\chi\over (2r)^3}.
\eqno (5.25)
$$
The right-hand side of this inequality is obviously less than 1 under conditions of Theorem 5.1.
Inequality (5.2) is proved.

\subsection{Relations and estimates for $T^{(r)}_{2s}(x)$}

Using (5.4) with $a=b=c=d=x$, we get the following
recurrence for the diagonal term $T^{(r)}_{2s}(x)$;
$$
T^{(r)}_{2s}(x) 
\le {1\over 4} \left( M*T^{(r)}(x)\right)_{2s-2} +
{1\over 4} \left( U(x)*D^{(r)}(x)\right)_{2s-2}
+{1\over 4} T^{(r+1)}_{2s-2}(x)
$$
$$
+{1\over 4} \left(T^{(2)}_{2s-2}(x)*D^{(r-1)}\right)_{2s-2}
+{1\over 4n} \left(U'(x)*U(x)*D^{(r-2)}\right)_{2s-2}
$$
$$
+{s\over n} \left(U(x)*D^{(r-1)}(x)\right)_{2s-2}
+ {s\over 2n} T^{(r)}_{2s-2}(x)
$$
$$
{r-2\over 4n^2} \left( U''*D^{(r-2)}\right)_{2s-2}
+{r-2\over 4n^2} \cdot {(2s+2)(2s+1)\over 2} \, T^{(r-1)}_{2s-2}(x,y)
\eqno (5.26)
$$
with the obvious initial condition $T^{(2)}_2 (x)= 1/(4n)$.

The right-hand side of (5.26) contains more terms than
those of the relations for the non-crossing terms $P^{(r)}_{2s}$ and 
the crossing term $Q^{(r)}_{2s}$.
Therefore the total number of terms to consider raises up to 16.
However, the estimates repeat in the most part
the estimates performed to prove (5.2). This is because
the second and the eights terms of the right-hand side of (5.21)
that are absent in (5.14) but present in (5.5) are of the
order smaller than the leading terms of the right-hand side of (5.21).
That is why the diagonal term $T^{(r)}_{2s}(x)$ is bounded by the same expression (5.3)
as the non-crossing term $Q^{(r)}_{2s}$ (5.2).

\vskip 0.5cm
We do not present the detailed proof of (5.3) because it is very similar to that of the proof of (5.2)
and uses the same formulas of Section 7.
Indeed, when estimating  $T^{(2r)}_{2s}$, we conclude from  relation (5.25)  and 
expressions (5.3) that the leading term and the negative part $ - \CX_4 - \CY_4$ are given 
by the corresponding terms of the right-hand side of relation (5.17). Then it is not hard to see
that the "extra" terms coming from the right-hand side of (5.26) with respect to (5.14)
add the terms
$$
{\chi \over 24} + {h^2\chi^2\over 3r(3r-1)(3r-2)}
\quad {\hbox{and }} \quad 
{h\chi\over 3} + { h\chi^3\over 12 (3r-1)(3r-2)}
$$
to the right-hand sides of (5.20) and (5.21), respectively. Certainly, this does not alter much the result of the sum
that is still  strictly less than 1 under conditions of Theorem 5.1.
 
\vskip 0.3cm
Regarding $T^{(2r+1)}_{2s}$, we see that the leading term and the negative contributions $\CX_5$
and $\CY_5$ are exactly the same as the corresponding terms of the right-hand side of 
(5.23). The "extra" terms of (5.26) produce then the 
additional terms 
$$
{4\chi\over (4r+1)(4r+3)}\quad {\hbox{and}} \quad {\chi^3\over 32} + {\chi\over 8(3r)^3} +
{3h \chi^3\over 8(3r)^3}
$$
for the right-hand sides of (5.24) and (5.25), respectively.
Certainly, this does not alter the result of sums that are 
strictly less than 1 under conditions of Theorem 5.1.

\section{Proof of Theorem 2.1}

Regarding  variable $R_n(s',s'')$  (4.3), we apply the result (4.6) of Theorem 4.1 to the factors of the first term 
$R^{(1)}_n(s',s'')$ and write that
$$
R^{(1)}_n(\bar s',\bar s'') \le n^2 \left[ \left( \vp(\t) + h  \t^2 n^{-2} (1-\t)^{-5/2} \right)^2\right]_{\bar s'}
\cdot \left[ \left( \vp(\t) + h  \t^2 n^{-2} (1-\t)^{-5/2} \right)^2\right]_{\bar s''},
$$
where $\bar s = s-1$. Repeating computations of (4.19),
we get inequality
$$
R^{(1)}_n(\bar s',\bar s'') \le 16 n^2  \left[  \vp(\t) + h  \t^2 n^{-2} (1-\t)^{-5/2} \right]_{\bar s'}
\cdot \left[  \vp(\t) + h  \t^2 n^{-2} (1-\t)^{-5/2} \right]_{\bar s''}
$$
$$
\le  16n^2\,  m_{\bar s'}\,  m_{\bar s''} \left( 1 + {h (\bar s')^3\over n^2}\right)  \left( 1 + {h (\bar s'')^3\over n^2}\right).
$$
It follows from the expression for $m_s$ (4.5) that 
$m_s = (\pi s^3)^{-1/2}(1+o(1))$ as $s\to\infty$. Then in the limit (2.5) we have the bound
$$
\limsup_{n\to\infty} R^{(1)}_n(\bar s',\bar s'') \le 16 { (1+h \chi')(1+h \chi'')\over \pi \sqrt{ \chi'\, \chi''}}.
\eqno (6.1)
$$
\vskip 0.5cm

Let us pass to the last term of the sum (4.3). Using the results of Theorem 5.1, we can write that
$$
R^{(4)}_{n}(\bar s',\bar s'') \le
\sum_{x,y=1}^n P^{(2)}_{2s'-2}(x,y) \, P^{(2)}_{2s''-2}(x,y) + \sum_{x=1}^n T^{(2)}_{2s'-2}(x)\, T^{(2)}_{2s''-2}(x)
$$
$$
\le 
n^2 \, \left({24s'C\over n^2} \right) \, \left({24s''C\over n^2} \right)
+ n \left( {6s'C\over n} \, m_{s'-1}\right) \, \left( {6s'C\over n} \, m_{ s''-1}\right). 
$$
Taking into account the asymptotic expression fro $m_s$,
we get  in the limit (2.5)
$$
R^{(4)}_{n}(\bar s',\bar s'') \le
\left( 24 C\over n^{2/3}\right)^2 \, (\chi' \chi'')^{1/3} (1+o(1))+
{ 36C^2\over n^{5/3}  \left( \chi'\, \chi''\right)^{1/6}} (1+o(1)).
\eqno (6.2)
$$
Similar computations show that
$$
R^{(2)}_n(s',s'') + R^{(3)}_n(s',s'') \le 
{96C \over n^{1/3}} \left( {(\chi'')^{1/3} (1+h\chi'') \over (\chi')^{1/2} }+ 
 {(\chi')^{1/3} (1+h\chi') \over (\chi'')^{1/2}}\right).
 \eqno (6.3)
 $$

Remembering the factor $V_4/n^2$ of (4.3), we see that 
$\hat \Sigma_n(s',s'')$ (3.3) vanishes in the limit (2.5) as $n\to\infty$.

\vskip 0.3cm 
Let us consider the variable $S_n(s',s'')$ (3.7) and its representation in four terms similar to (4.3). 
It follows from the results of Section 4
that the terms $S^{(k)}_n(s',s'')$, $k=1,2,3$ that contain factors $U_\a(x,y) U_\b(x,y)$ are equal to zero.
Then only the term 
$$
S^{(4)}_n(s',s'') = \sum_{x,y=1}^n
\left( \sum_{\a_1+\b_1 = 2s'} \E\left\{ (A^{\a_1}_{xy})^\circ \,
(A^{\b_1}_{xy})^\circ\right\}\right)
\left( \sum_{\a_2+\b_2 = 2s''} \E\left\{ (A^{\a_2}_{xy})^\circ \,
(A^{\b_2}_{xy})^\circ\right\}\right),
$$
gives a non-zero contribution to $S_n(s',s'')$.  Repeating computations of Subsection 5.2, it is not hard to show that
$S^{(4)}_n(s',s'') = o(1)$  in the limit (2.5) as $n\to\infty$.
This completes the proof of Theorem 2.1.

\section{Auxiliary relations}

For completeness, let us refer to some of the  equalities and identities of \cite{K2} that we use
in the present paper.
The first equality is a consequence of the integration by parts formula
applied to the normal  random variable $\xi \sim \CN(0,v^2)$ that is 
$\E \xi f(\xi) = v^2 \E f'(\xi)$, where $f(x)$ is a non-random function such that 
corresponding mathematical expectations exist.  Then for the random matrix $A$
from GUE we get, in particular, relation
$$
\E A_{xy} \left( A^k\right)_{uv} = {1\over 4n} \, \sum_{j=0}^{k-1} \E \left\{ (A^j)_{uy}\, (A^{k-1-j})_{xv}\right\}.
\eqno (7.1)
$$

\vskip 0.1cm
The second identity relates the generating functions $(1-\t)^{-r-1/2}$ with the moments $m_s$. 
It is not hard to obtain that 
$$
\left[ {1\over (1-\t)^{r+1/2}}\right]_s = r{  {2r+2s\choose 2s} \over  { r+s\choose s+1}} \, m_s,
\eqno (7.2)
$$
or in equivalent form,
$$
\left[ {1\over (1-\t)^{r+1/2}}\right]_s = {1\over 2^{2s}\, s!} \cdot { (2r+2s)!\over (r+s)!} \cdot 
{r!\over (2r)!}.
\eqno (7.3)
$$
Two particular cases are important:
$$
{(2s+1)(2s+2)\over 2} \, m_s = \left[ {1\over (1-\t)^{3/2}}\right]_s
\eqno (7.4)
$$
and 
$$
{(2s+1)(2s+2)(2s+3)\over 3!} \, m_s = \left[ {1\over (1-\t)^{5/2}}\right]_s.
\eqno (7.5)
$$

We also use the equality
$$
 \left[ {1\over (1-\t)^{k+1}}\right]_s = { (s+1)(s+2)\cdots (s+k)\over k!} .
\eqno (7.6)
$$

\end{document}